\documentclass[12pt]{article}
\usepackage{newtxtext,newtxmath}
\usepackage{braket}

\usepackage{xcolor}

\usepackage{graphicx}

\usepackage{bm}
\usepackage[letterpaper,margin=1in]{geometry}

\renewenvironment{abstract}
	{\quotation}
	{\endquotation}

\date{}

\makeatletter
\renewcommand{\fnum@figure}{\textbf{Figure \thefigure}}
\renewcommand{\fnum@table}{\textbf{Table \thetable}}
\makeatother

\usepackage{scicite}

\usepackage{url}

\newcommand{\cm}{\ensuremath{\,\mathrm{cm^{-1}}}}
\newcommand{\wpi}{W-\ensuremath{\pi}\ }

\def\scititle{
	Electron dynamics mediate the water-carbon $\pi$ bond
}

\title{\bfseries \boldmath \scititle}

\author{
	N.~LeMessurier$^{1\dagger}$,
	E.~J.~Katz$^{2,3\dagger}$,
    R.~Pant$^1$,
    S.~E.~Ganley$^{1,2}$,
    H.~Salzmann$^{1,2}$,\and
    L.~M.~McCaslin$^{4}$,
    J.~M.~Weber$^{1,2\ast}$,
    J.~D.~Eaves$^{1\ast \ast}$\\
	\small$^{1}$Department of Chemistry, University of Colorado, Boulder CO 80309, USA \and
	\small$^{2}$JILA, University of Colorado, Boulder, Boulder CO 80309, USA \and
    \small${^3}$ Department of Physics, University of Colorado, Boulder CO 80309, USA \and
    \small$^{4}$Sandia National Laboratories, Livermore, CA 94550, USA \and
	\small$^{\ast}$Corresponding author. Email: weberjm@jila.edu \and
    \small$^{\ast \ast}$Corresponding author. Email: joel.eaves@colorado.edu \and
	\small$^{\dagger}$These authors contributed equally to this work.
}

\begin{document} 

\maketitle

\begin{abstract} \bfseries \boldmath
The intermolecular interaction between a water molecule and the electrons in aromatic $\pi$ systems, the water-$\pi$ bond, lies at the heart of many chemical processes, yet its properties remain challenging to measure experimentally and model computationally. Infrared spectroscopy of pyrene anions hydrated by a single water molecule reveals vibrational and electronic motions that are often hidden in condensed phase measurements. New machine-learning approaches to potentials and dipole moments show that the electron dynamics of the aromatic $\pi$ cloud quench some signals from water's vibrations and amplify others. The observed interplay between electronic and vibrational motions has general implications for modeling spectra and intermolecular interactions between water and aromatic systems in clusters, solutions, and at interfaces.
\end{abstract}

\newpage

\noindent
Weak intermolecular forces between water and the electrons in the $\pi$-bonds of aromatic molecules determine the chemical and physical properties of many substances. For instance, the water-carbon $\pi$ bond (\wpi bond) competes with stronger hydrogen bonds and ionic interactions to stabilize proteins and nucleic acids in solution \cite{AminoAcidsRef}. It also mediates the hydrophobic effect and electrochemistry at interfaces, such as graphene, \cite{Sacchi31122023,MOJIRI2019133971} and is critical for ice nucleation on carbonaceous dust \cite{QUIRICO201632}.

The bond strength of the \wpi interaction is comparable to that of a weak hydrogen bond, and is sometimes referred to as a $\pi$-hydrogen bond\cite{feller1999strength-bdb, gierszal2011-hydrogen-71f}. While it is paramount in many chemical contexts, studying the \wpi interaction experimentally is challenging, complicating efforts to develop accurate computational models \cite{suzuki1992benzene-d1b}. In the solution phase, signals from molecules engaged in \wpi bonds are difficult to detect\cite{gierszal2011-hydrogen-71f}. In particular, the infrared (IR) spectrum of the OH stretching region of water directly probes intermolecular structure and the dynamics of charge fluctuations that occur on timescales of $10 - 1,000$ femtoseconds (fs), but the broad and intense background of the bulk phase typically overwhelms the signatures of molecules in \wpi bonds \cite{Auer1}. 

In computational studies, an accurate and computationally efficient model of the \wpi interaction is crucial.\cite{schravendijk2005from-7e5} For decades, researchers have used empirical force fields to model intermolecular forces in aqueous chemistry, often with great success \cite{mackerell1998all-atom-e5f}. Such models employ fixed atomic charges to efficiently parameterize polar and ionic interactions\cite{lopes2004modeling-a0d}, but the \wpi bond is composed of more complex noncovalent forces that may not necessarily lend themselves to parameterization by fixed charges\cite{allesch2008first-651}. The \wpi bond remains a steep challenge for molecular modeling. 

In recent years, there has been an explosion in machine learning (ML) techniques that minimize user bias, like point charge assignments, in fitting potentials\cite{behler2014representing-f36,bartk2017machine-0f1}. These methods offer new possibilities for representing a weak intermolecular interaction like the \wpi bond, without introducing atomic charges\cite{schran2021machine-c45}. Like empirical force fields, ML potentials can enable the computation of properties such as correlation functions, phase boundaries, and chemical reaction pathways with statistical significance and at accuracies limited only by the electronic structure calculations used to fit the potential \cite{kovcs2023evaluation-a40,bhatia2025leveraging-07f}. The tradeoff is that the ML framework is more data-based than physics-based, potentially sacrificing interpretability for accuracy, as ML potentials can have orders of magnitude more parameters than many empirical potentials do \cite{batatia2022mace-4b7}. While ML potentials achieve this by eschewing parameterizations such as point charges that can facilitate chemical intuition and aid model extrapolation, the intermolecular interactions they describe are more faithful to the full quantum nature of the electron. From this perspective, ML potentials implicitly encode realistic electronic dynamics that experiments can probe by measuring fluctuations in the dipole moment. 

\subsection*{Hydrated anionic clusters expose the water-$\pi$ bond}
In this work, we use mass-selected cluster ions consisting of a single pyrene molecule and a single water molecule to study \wpi interactions. In these clusters, the polycyclic aromatic hydrocarbon (PAH) molecule is a negative ion whose excess charge strengthens the \wpi bond, making it the dominant attractive intermolecular force \cite{knurr1,Lemessurier1,Schiedt1}.

Mass-selected clusters are model systems for studying the \wpi bond because they circumvent many of the challenges that complicate experiments in the condensed phase \cite{robertson1,vernon1}. First, the vibrational resonances of the OH stretching modes are sharper than they are in the condensed phase, revealing dynamical line-broadenings that are otherwise obscured.  Second, water molecules interacting with aromatic systems in solutions or at interfaces are minority species whose spectroscopic contributions are often too small to resolve. Finally, the number of water molecules interacting with a given molecular species at any given time fluctuates, leading to problems of speciation. In contrast, the \wpi bond is the only intermolecular interaction in a singly hydrated PAH cluster \cite{Lemessurier1,knurr1,Heinrich1}. 

We measure the cluster's IR spectrum using photodissociation action spectroscopy \cite{vernon1,Farrar_Saunders_1988,Lemessurier1} (see Supplementary Materials for experimental details). The target clusters are tagged with loosely bound argon (Ar) atoms and irradiated with tunable IR radiation. A PAH$^- \cdot$H$_2$O$\cdot$Ar$_\mathrm{n}$ cluster that absorbs a photon will lose the Ar tags via vibrational predissociation. The signal of the resulting fragment ions PAH$^- \cdot$H$_2$O (after secondary mass analysis) as a function of the IR frequency is a surrogate for the IR spectrum of the cluster. The presence of the Ar atoms also limits the temperature in the experiments to tens of K.

We also employ deuteration of the water molecule through all of its forms, H$_2$O, HOD, and D$_2$O, to obtain additional information on the vibrational dynamics of the cluster. In the Born-Oppenheimer approximation, a central tenet of molecular structure theory, both the intermolecular potential and the electron dynamics are invariant under isotopic substitution \cite{eltareb2025understanding-12a}. In HOD the isotope effects simplify the character of the vibrations. The molecule has one high-frequency OH local stretching mode and one local OD stretching mode. In D$_2$O, the isotope shift moves the symmetric and antisymmetric modes to lower frequencies. While the forces on the atoms are the same as in H$_2$O, the heavier deuteron moves more slowly than the protons, leading to slower dynamics that manifest as line narrowing. Because deuteration does not affect electron behavior, \cite{eltareb2025understanding-12a} comparison between spectra with and without deuteration can help differentiate electron and nuclear dynamics. 

\subsection*{Predictions of Intermolecular Statics and Dynamics}
\begin{figure}[htp] 
	\centering
	\includegraphics[width=140mm]{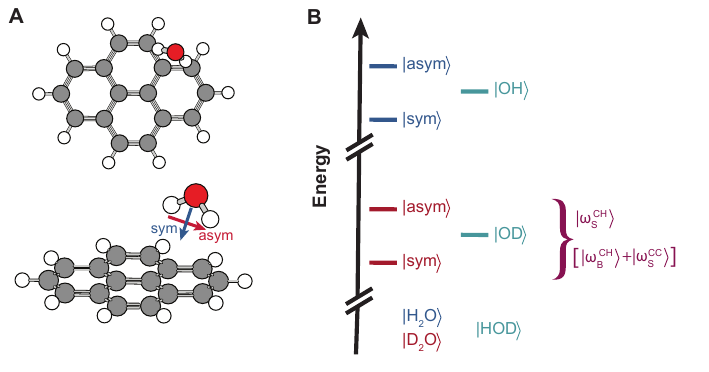} 
	\caption{\textbf{Structure and energy level diagram for the vibrations of the anionic pyrene monohydrate cluster in its deuterated forms.} A typical configuration of the pyrene monohydrate anion cluster and the directions of the symmetric (sym) and antisymmetric (asym) transition dipole moments (A) along with relative vibrational energies for the water and pyrene moieties (B). Both H$_2$O and D$_2$O have symmetric and antisymmetric stretching vibrations, depicted in dark blue and red, respectively. Deuteration in HOD introduces an isotope effect that breaks the local symmetry of the water molecule and separates vibrations into two local modes,  $\ket{\mathrm{OH}}$ and $\ket{\mathrm{OD}}$, depicted in teal. The OH resonances occur above $\approx$ 3500 cm$^{-1}$ and are vibrations of the water molecule. The transitions at lower frequencies overlap the spectral region of the CH stretching fundamentals, $\ket{\omega_s^{CH}}$, and the CC stretching/CH bending combination bands ($\ket{\omega_B^{CH}}+\ket{\omega_s^{CC}}$ of the pyrene moiety \cite{Heinrich2}, shown in purple.  This region of the spectrum is thus more complicated to interpret than the resonances from the OH vibrations. More details appear in the Supplementary Materials.}
	\label{fig:intro_molec} 
\end{figure}

Fig. \ref{fig:intro_molec}A depicts the pyrene monohydrate cluster anions studied in this work, showing a typical intermolecular configuration in which the OH groups form \wpi bonds. The energy level diagram of the OH and OD stretching vibrations in the water molecules appears in Fig. \ref{fig:intro_molec}B alongside the vibrational transitions of the pyrene molecule. For H$_2$O, the frequencies are well-separated from transitions of the pyrene moiety, but in the OD stretching region of the D$_2$O and HOD molecules, there can be some spectral overlap near 2600 \cm \cite{Heinrich2}. The stabilizing water-$\pi$ interaction red-shifts the OH stretching modes of the water molecule relative to the gas phase, and the dynamics of the water molecule exploring the configurational space of its complex with the PAH lead to line broadening, discussed below. 

Fig. \ref{fig:free_energies} highlights some distinguishing characteristics between the empirical (A, B) and ML (C, D) potential energy functions. To parameterize an empirical model, we follow standard procedures and fit a model that provides reasonable agreement with many features of the PAH-H$_2$O spectra and the absorption spectra of liquid water, from the terahertz to the IR region \cite{Lemessurier1}. The intramolecular part of the water potential comes from a gas phase model that, in mixed quantum-classical simulations, yields accurate absorption and photon echo spectra for the OH stretching region of HOD in liquid D$_2$O \cite{reimers_local_1984}. The intramolecular interactions between the PAH and the water molecule combine the TIP4P/2005 water model,\cite{abascal_general_2005} a popular model for water at ambient conditions, with the DREIDING force field,\cite{mayo_dreiding_1990} which describes noncovalent interactions between the water and PAH molecule and the atomic intramolecular interactions in the PAH. While the TIP4P/2005 model specifies the water molecule's charges, one must calculate the point charges on the PAH. The point charges in the empirical potential, assigned using the Merz-Singh-Kollman ESP method\cite{singh_approach_1984,besler_atomic_1990} that reproduces the electrostatic potential at distances far from the pyrene moiety, appear in Fig. \ref{fig:free_energies}A. 

In contrast to the empirical potential, the ML potential does not use point-charge assignments to parameterize intermolecular interactions. The diffuse electron density depicted in Fig. \ref{fig:free_energies}C is a stark contrast to the point charge distribution in Fig. \ref{fig:free_energies}A. The ML model is likely more accurate at short intermolecular separations where the accuracy of the multipole approximation breaks down\cite{stocco2025electric-field-2af}. For the ML model, we employ the MACE potential because it is equivariant, data-efficient, and transferable \cite{batatia2022mace-4b7,batatia2022design-dc7}. The fitting procedure, including the construction of the training sets, appears in the Supplementary Materials. 

Before turning to the dynamics from molecular dynamics (MD) simulations, we analyze and compare the free energy landscapes computed with the empirical force field and the DFT-trained machine-learned interatomic potentials (MLIP). All computational data come from MD simulations, with trajectories computed using either the empirical or ML potential. At the temperatures relevant for the present work ($\approx$ 75 K), the water molecules are mobile enough to explore the surface of the pyrene molecule. The free energy surface quantifies the differences in intermolecular structures $F(x,y)=-k_B T \ln(P(x,y))$, where $P(x,y)$ is the joint probability distribution for finding the oxygen atom of the water molecule at position $x$ and $y$. The free energy in Fig. \ref{fig:free_energies}B shows two minima centered on the pyrene molecule where the water is most likely to lie. Two valleys connect the minima with a barrier of a few $k_B T$. The free energy for the ML potential in Fig. \ref{fig:free_energies}D is markedly different. The two basins are present in the free energy surfaces for both potentials, but in the ML potential, the free energy barrier separating the basins is significantly lower, and the basins are wider. The shape of the free energy in the ML potential resembles a butterfly, with additional local minima at the points of the butterfly wings, off the carbon frame of the PAH (asterisk marks, Fig. \ref{fig:free_energies}D). Those minima are absent in the free energy from the empirical potential. The low barriers separating basins imply that the water molecules are much more mobile over the face of the anionic pyrene in the ML potential than in the empirical one.

\subsection*{Nuclear and Electronic Fluctuations of the Dipole Moment}
\begin{figure}[htp] 
	\centering
	\includegraphics[width=160mm]{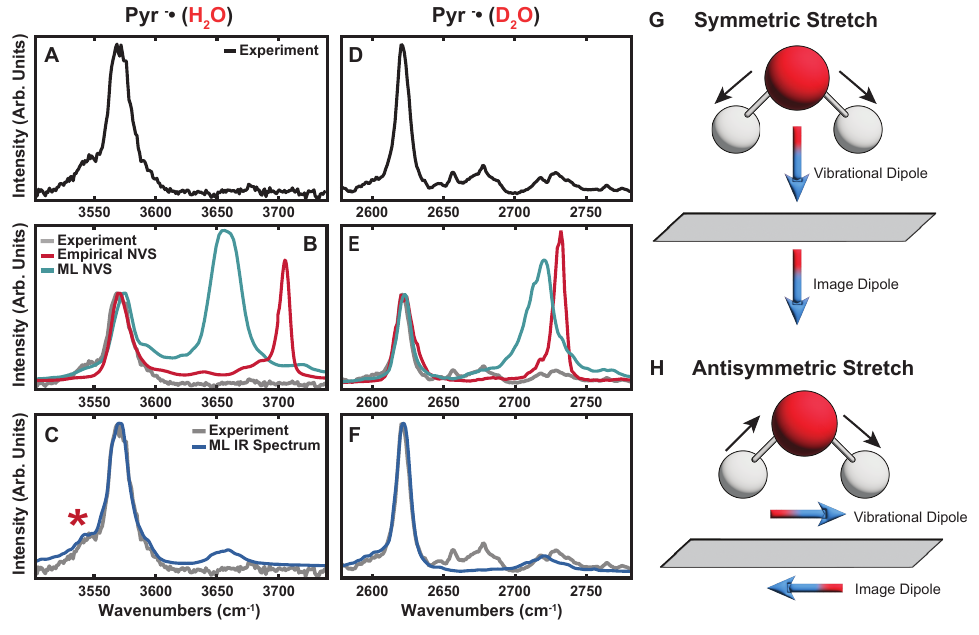} 
	\caption{\textbf{Spectra of the OH and OD stretching regions for Pyr$^-\cdot$H$_2$O and Pyr$^-\cdot$D$_2$O compared to predictions from the empirical and machine-learned models highlight the relative roles of electronic and vibrational dynamics.} (Top left panels) Experimental IR spectra of Pyr$^-\cdot$H$_2$O$\cdot$Ar$_2$ (A, taken from ref. \cite{Lemessurier1}) and Pyr$^-\cdot$D$_2$O$\cdot$Ar$_2$ (D). (Middle left panels) Nuclear vibrational spectra (NVS) calculated at 75 K with the empirical potential (red) and the ML potential (teal) show nuclear dynamics but neglect electron dynamics. Both potentials show an intense antisymmetric stretching peak that does not appear in the experimental data (repeated in gray). (Bottom) The IR spectra calculated from ML dynamics using the fluctuations of the total dipole moment, inferred from a machine-learned method, (blue) match the experimental data more closely. Those simulations reproduce the suppression of the intensity of the antisymmetric stretching fundamental relative to the symmetric one in both normal and heavy water (C, F).  The electrons in the $\pi$ orbitals of the Pyr$^-$ molecule follow the nuclear vibrations, forming an image radiation dipole that quenches the signal from the antisymmetric vibration, lying mostly in the pyrene plane, while amplifying the symmetric vibration, which lies mostly normal to the pyrene plane (G,H).}
	\label{fig:ir_h2o_d2o} 
\end{figure}

Experimental IR spectra appear in Figs. \ref{fig:ir_h2o_d2o}A and D for the pure isotopologues of the water molecule in both the OD and OH stretching regions. The spectra of the complexes with H$_2$O \cite{Lemessurier1} (Fig. \ref{fig:ir_h2o_d2o}A) and D$_2$O (Fig. \ref{fig:ir_h2o_d2o}D) share some similarities. Indeed, the D$_2$O spectrum nearly duplicates the H$_2$O spectrum, only at lower frequencies and with a slightly narrower lineshape, and there are some additional weak features originating from the spectrum of the pyrene anion (see Supplementary Materials). 

Recent breakthroughs have enabled ML algorithms to infer the total dipole moment from a computed trajectory, enabling a more accurate representation of the fluctuating dipoles, including nuclear and electronic contributions, thus allowing electron dynamics that are suppressed in conventional models to manifest in spectra\cite{gastegger2017machine-569}. In contrast to more conventional calculations that compute only the vibrational contribution to the dipole moment, inference methods account for both electronic and vibrational contributions\cite{bhatia2025leveraging-07f}. We refer to the spectra computed using only the vibrations as nuclear vibrational spectra (NVS), and the spectra computed from the inferred dipole moment as IR spectra. Importantly, both spectra are computed from the same molecular trajectories. By systematically analyzing those spectra at various levels of deuteration, we isolate the roles of molecular structure, vibrational motion, and electron dynamics in the measured IR spectra of pyrene monohydrates. 

To interpret and assign the experimental IR spectra, we compute two types of spectra using the time-correlation function formalism. The spectral intensities are $I(\omega)=Q(\omega)C(\omega)$, where $C(\omega)$ is the Fourier transform of the dipole-dipole time correlation function, $C(\omega)=\int_{-\infty}^\infty dt e^{-i\omega t} \langle {\bm \mu}(t)\cdot{\bm \mu}(0)\rangle $, and $Q(\omega)$ is a function that accounts for detailed balance and macroscopic electrodynamics \cite{zwanzig2001nonequilibrium-8c1}. To isolate the vibrational contributions from the electronic ones, we compute the NVS using only the nuclear contributions to the dipole moment. The NVS $I_{vib}(\omega)$ comes from the associated vibrational correlation function $C_{vib}(\omega)$, and is equivalent to the IR spectrum when all atomic charges are fixed. To calculate the IR spectrum $I_{IR}(\omega)$, which contains both electronic and vibrational contributions, we infer the total dipole moment as a function of time along trajectories from the ML potential. The resulting dipole moment time correlation function is $C_{IR}(\omega)$. 

\begin{figure}[htp]
	\centering
	\includegraphics[width=120mm]{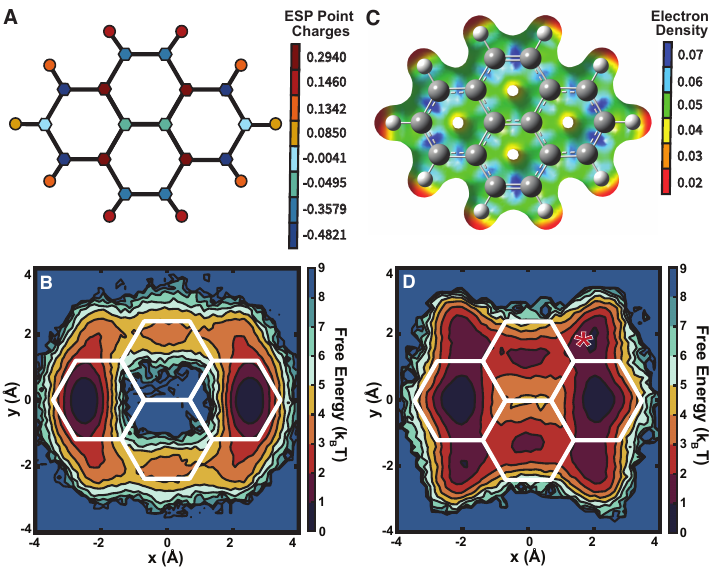} 
	\caption{\textbf{Free energy landscapes of the water molecule moving across the surface of the pyrene anion highlight the differences that the empirical model and the machine-learned one predict for intermolecular structure.} The empirical model assigns point charges to atoms of Pyr$^-$ according to the Merz-Singh-Kollman ESP technique (A) and corresponding free energy derived probability density at 75 K for the oxygen of the water molecule to lie in the plane of the Pyr$^-$ molecule (white frame) (B). The electron density from electronic structure calculations (C) is shaped and more diffuse than the skeletal charge distribution in (A), which can impact interactions at short distances. The free energy surface from the machine-learned potential is markedly different from that in (B), showing four new basins at the points of the butterfly shape, marked with a red asterisk in the top right quadrant. All electronic structure calculations, including those that generate the training data for the machine-learned potential, use density functional theory with the $\omega$B97X-D/def2-TZVPP  method and basis set.}
	\label{fig:free_energies} 
\end{figure}

The nuclear vibrational spectra $I_{vib}(\omega)$, accounting for nuclear fluctuations in the dipole moment, from both empirical and ML potentials exhibit a symmetric stretching feature with linewidth and lineshape similar to those observed experimentally. The structures at the points of the butterfly wings that were missing in the empirical potential (red asterisk, Fig. \ref{fig:free_energies}D) appear as a shoulder on the red side of the symmetric stretching resonance in both H$_2$O and D$_2$O (red asterisk, Fig. \ref{fig:ir_h2o_d2o}). As those structures were missing in the empirical potential, they do not appear in the NVS from the empirical potential.

The most striking difference between the measured IR spectra and the computed NVS for both H$_2$O and D$_2$O lies in the intensity of the antisymmetric stretching vibration. For water in the gas phase, the IR intensity of the antisymmetric stretching vibration is significantly larger than the symmetric stretching transition \cite{galabov2002high-b9d}. In contrast, the experimental IR spectra of the analogous antisymmetric mode in the pyrene-water cluster are much less intense than the symmetric one. While the experimental IR spectra show a strong suppression of the antisymmetric mode, both the empirical and ML potentials predict an intense antisymmetric resonance. While the two potentials give somewhat different widths and positions of the antisymmetric vibrations, they both predict that the antisymmetric vibration is at least as intense as the symmetric one.

The electron dynamics come from a MACE model trained to infer the total dipole moment ${\bm \mu}$ from atomic positions ${\bm R}$, such that ${\bm \mu(t)} = {\bm \mu}({\bm R}(t))$. With the trajectories ${\bm R}(t)$ from the ML potential, the IR spectra recover the suppression of the antisymmetric peak in H$_2$O and D$_2$O (Fig. \ref{fig:ir_h2o_d2o}C, F). The computed IR spectra from the total dipole fluctuations, including electronic and vibrational contributions, match the experimentally measured spectra quantitatively, including the asymmetric lineshape of the symmetric stretching feature and the relative line-narrowing between H$_2$O and D$_2$O.

The analysis using the ML models shows that electron dynamics are solely responsible for the suppression of the antisymmetric stretching signature. To understand this effect, consider how the electrons in pyrene respond to the vibrations of the water molecule. The water molecule induces a dipole in the electron cloud of the pyrene anion that behaves as an image dipole (Figs. \ref{fig:ir_h2o_d2o}G, H), oscillating in response to the water molecule's vibrations. On average, the water molecule donates both of its protons into the \wpi bond, so that the transition dipole of the symmetric stretching vibration points normal to the pyrene plane and the antisymmetric transition dipole lies parallel to that plane. Because the image dipole lies in the mirror plane, it amplifies the symmetric mode and mutes the antisymmetric one (Figs. \ref{fig:ir_h2o_d2o}G, H). Electron dynamics from vibration-induced charge transfer between the two molecules may also be present, but an elementary calculation shows that they are significantly smaller than the image dipole effects (see Supplementary Materials). Charge transfer dynamics do not comprehensively explain the relative enhancement and suppression of resonances in the spectrum for all isotopes of water, while the induced image dipole does.

It is unusual for the electrodynamics of one molecule (pyrene) to dictate the IR spectrum of another (water). While the transition dipoles from static electronic structure theory of low-energy structures also predict a small transition dipole moment for the antisymmetric mode, they do not reveal the mechanism that explains the suppression of the IR intensity in this mode. It is not straightforward in adiabatic electronic structure calculations to separate vibrational fluctuations of the water molecule from the electronic motions of the cluster. The electrodynamic mechanism we propose is familiar in radiation theory on much larger length scales, but may be unexpectedly commonplace in molecules. The Chance-Prock-Silbey theory\cite{chance2016advances-f34}, for example, invokes a similar picture to ours when describing the fluorescence of a molecule near a metallic surface, borrowing ideas from antenna communication near ground planes \cite{adekola2019effects-de3}. In both cases, whether the image dipoles reinforce or destroy radiating fields depends on the orientation of the radiating dipole relative to the surface. 

\begin{figure}[htp] 
	\centering
	\includegraphics[width=160mm]{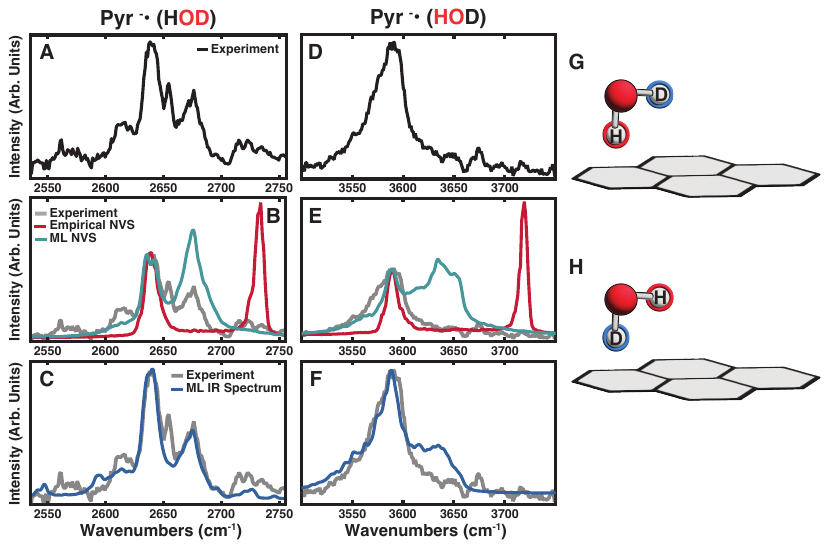} 
	\caption{\textbf{Spectra of the OH and OD stretching regions for Pyr$^-\cdot$HOD from experiment and computation.} (Top) The experimental IR spectra of Pyr$^-\cdot$HOD$\cdot$Ar$_2$ in the OD stretching (A) and OH stretching regions (D). Some of the vibrational signatures in the OD stretching region originate from vibrational levels of the pyrene anion (see Supplementary Materials). (Middle) Nuclear vibrational spectra (NVS) calculated at 75 K (B, E) with the empirical potential (red) and ML potential (teal) neglect electron dynamics and have a large peak for the blue component of each doublet. These transitions are not found in the experiment (gray). (Bottom) The IR spectra calculated from ML dynamics at 75 K with an ML-predicted dipole (blue) match the experimental data more closely (C, F). The doublet structure emerges from different average orientations of the OH and OD groups (G,H). The tilted orientation of the HOD molecule is exaggerated here for illustration.}
	\label{fig:ir_hod} 
\end{figure}

The IR spectra for HOD appear in Fig. \ref{fig:ir_hod}. The mass difference between the proton and deuteron splits the antisymmetric and symmetric stretching modes into two local modes---one OH oscillation at 3590 \cm and one OD oscillation at 2640 \cm. The dynamics of the water molecule on the surface of the carbon frame lead to transient structures that can have one proton or deuteron engaged in a \wpi bond and one disengaged from it (Fig. \ref{fig:ir_hod}G, H). These transient structures split the OH and OD resonances further into two doublets, where the bound OH or OD oscillator is red-shifted relative to the free one. The mode of the bound oscillator carries the character of the symmetric stretching mode, while the free oscillator maps to the antisymmetric mode. In our electrodynamic interpretation, the transition dipole moment of the free oscillator is largely parallel to the pyrene plane, while that of the bound oscillator is mostly normal to it. Thus, the blue component of each doublet should be suppressed relative to the red one. The calculated NVS and IR spectra show that trend. In HOD, both ML and empirical potentials predict intense resonances for the antisymmetric stretching vibrations, though the peak and width of the resonance differ more substantially between the two potentials in HOD than they do in D$_2$O or H$_2$O. The electron dynamics inferred in the ML model quench the antisymmetric peak (Fig. \ref{fig:ir_hod}C, F), making the calculated IR spectra much more similar to the experimental data, though the agreement between theoretical predictions and experimental results is better for the OD oscillator than the OH one. 

Anionic pyrene monohydrate clusters are a model system for studying the \wpi bond using vibrational spectroscopy and highly accurate theoretical models, which were computationally infeasible to implement before some of the ML methods employed in this paper were available. The system reveals vibrational dynamics in the \wpi bond that condensed phase systems obscure, while a series of experiments employing deuteration allows for the rational interpretation of the ML models in terms of the electron and nuclear dynamics they encode. Decades of data from MD simulations show that fixed charges are often an excellent approximation to the electrostatic interactions between a polar molecule, like water, and ionic species\cite{lopes2004modeling-a0d,joung2009molecular-797,banerjee2018rotational-0fe}. Our results provide an important counterpoint, where the $\pi$ electrons on the pyrene moiety move in concert with the vibrational oscillations of the water molecule, screening the antisymmetric stretching dipole in H$_2$O and D$_2$O and the blue component in the doublets of both hydride oscillations in HOD. The electrodynamic screening phenomenon is remarkably consistent across the series of deuterated clusters. While the atomic forces and vibrational spectra from the fixed charge models can be reasonably accurate, they cannot describe, even qualitatively, the intermolecular electrodynamic screening reported here. The analogy with image charge phenomena in macroscopic matter invokes these fundamental concepts on a molecular scale, showing that the dynamics of the $\pi$ electrons are a key property that must be taken into account when describing the \wpi bond. Those electronic fluctuations lead to a strengthening of the \wpi bond that fixed charge models cannot describe. While the effects may be subtle in solution phase chemistry, they are likely to be much more pronounced at the aqueous-carbon interface, manifest in the hydrophobicity and electrowetting of graphitic membranes \cite{ostrowski2014tunable-303, belyaeva2020wettability-186}.

\newpage

\subsection*{Acknowledgments}
We gratefully acknowledge support from the Department of Energy, Office of Basic Energy Sciences, under award number DE-SC0021387. This work utilized the Alpine high-performance computing resource at the University of Colorado Boulder. Alpine is jointly funded by the University of Colorado Boulder, the University of Colorado Anschutz, and Colorado State University and with support from NSF grants OAC-2201538 and OAC-2322260.

This material is based upon work supported by the U.S. Department of Energy, Office of Science, Office of Workforce Development for Teachers and Scientists, Office of Science Graduate Student Research (SCGSR) program. The SCGSR program is administered by the Oak Ridge Institute for Science and Education for the DOE under contract number DE-SC0014664.

L.M.M. was supported by Sandia National Laboratories through the Gas Phase Chemical Physics Program in the Division of Chemical Sciences, Geosciences and Biosciences, Office of Basic Energy Sciences (BES), United States Department of Energy (U.S. DOE). Sandia National Laboratories is a multi-mission laboratory managed and operated by the National Technology \& Engineering Solutions of Sandia, LLC (NTESS), a wholly owned subsidiary of Honeywell International, Inc., for the U.S. Department of Energy’s National Nuclear Security Administration (DOE/NNSA) under Contract DE-NA0003525. This written work is authored by an employee of NTESS. The employee and not NTESS owns the right, title, and interest in and to the written work and is responsible for its contents. Any subjective views or opinions that might be expressed in the written work do not necessarily represent the views of the U.S. Government. The publisher acknowledges that the U.S. Government retains a non-exclusive, paid-up, irrevocable, world-wide license to publish or reproduce the published form of this written work or allow others to do so for U.S. Government purposes. The U.S. DOE will provide public access to results of federally sponsored research in accordance with the U.S. DOE Public Access Plan. This research used resources of the National Energy Research Scientific Computing Center (NERSC), a Department of Energy User Facility using NERSC award BES-ERCAP 0032345.

{\bf Author Contributions}: NL and RP: Investigation (computational), Formal analysis.
EJK, SEG and HS: Investigation (experimental), Formal analysis.
JWM, LMM, and JDE: Conceptualization, Supervision.
JDE and JWM: Writing; produced the original draft.
All authors: Writing; review \& editing; approval of final manuscript.

{\bf Competing Interests}: The authors declare that they have no competing interests.

{\bf Data Availability}: All data needed to evaluate the conclusions in the paper are present in the manuscript and the Supplementary Materials. Additional data, code, and trained ML models will be made available upon reasonable request to the corresponding authors and will be deposited in a publicly accessible repository upon acceptance.

\bibliographystyle{sciencemag}
\bibliography{science_pyrene_monohydrate}

@article{AminoAcidsRef,
author = {Scheiner, Steve and Kar, Tapas and Pattanayak, Jayasree},
title = {Comparison of Various Types of Hydrogen Bonds Involving Aromatic Amino Acids},
journal = {Journal of the American Chemical Society},
volume = {124},
number = {44},
pages = {13257-13264},
year = {2002},
doi = {10.1021/ja027200q},
note ={PMID: 12405854},
URL = { https://doi.org/10.1021/ja027200q},
}

@article{Heinrich1,
author = {Salzmann, Heinrich and LeMessurier, Natalie and Eaves, Joel D. and Weber, J. Mathias},
title = {Formation of Water Networks on Anionic Perylene},
journal = {The Journal of Physical Chemistry A},
volume = {129},
number = {20},
pages = {4384-4393},
year = {2025},
doi = {10.1021/acs.jpca.5c00569},
note ={PMID: 40334062},
URL = { https://doi.org/10.1021/acs.jpca.5c00569},
}

@article{Lemessurier1,
author = {LeMessurier, Natalie and Salzmann, Heinrich and Leversee, River and Weber, J. Mathias and Eaves, Joel D.},
title = {Water–Hydrocarbon Interactions in Anionic Pyrene Monohydrate},
journal = {The Journal of Physical Chemistry B},
volume = {128},
number = {13},
pages = {3200-3210},
year = {2024},
doi = {10.1021/acs.jpcb.3c07777},
note ={PMID: 38526297},
URL = { https://doi.org/10.1021/acs.jpcb.3c07777},
}

@article{robertson1,
author = "Robertson, William H. and Johnson, Mark A.",
title = "Molecular Aspects of Halide Ion Hydration: The Cluster Approach", 
journal= "Annual Review of Physical Chemistry",
year = "2003",
volume = "54",
number = "Volume 54, 2003",
pages = "173-213",
doi =
"https://doi.org/10.1146/annurev.physchem.54.011002.103801",
url = "https://www.annualreviews.org/content/journals/10.1146/annurev.physchem.54.011002.103801",
publisher = "Annual Reviews",
issn = "1545-1593",
type = "Journal Article",
}

@article{Schiedt1,
author = {Schiedt, J. and Knott, W. J. and Le Barbu, K. and Schlag, E. W. and Weinkauf, R.},
title = {Microsolvation of similar-sized aromatic molecules: Photoelectron spectroscopy of bithiophene–, azulene–, and naphthalene–water anion clusters},
journal = {The Journal of Chemical Physics},
volume = {113},
number = {21},
pages = {9470-9478},
year = {2000},
month = {12},
issn = {0021-9606},
doi = {10.1063/1.1319874},
url = {https://doi.org/10.1063/1.1319874},
}

@article{vernon1,
author = {Vernon, M. F. and Krajnovich, D. J. and Kwok, H. S. and Lisy, J. M. and Shen, Y. R. and Lee, Y. T.},
title = {Infrared vibrational predissociation spectroscopy of water clusters by the crossed laser‐molecular beam technique},
journal = {The Journal of Chemical Physics},
volume = {77},
number = {1},
pages = {47-57},
year = {1982},
month = {07},
issn = {0021-9606},
doi = {10.1063/1.443631},
url = {https://doi.org/10.1063/1.443631},
}

@article{knurr1,
author = {Knurr, Benjamin J. and Adams, Christopher L. and Weber, J. Mathias},
title = {Infrared spectroscopy of hydrated naphthalene cluster anions},
journal = {The Journal of Chemical Physics},
volume = {137},
number = {10},
pages = {104303},
year = {2012},
month = {09},
issn = {0021-9606},
doi = {10.1063/1.4750371},
url = {https://doi.org/10.1063/1.4750371},
}

@article{Auer1,
author = {B. Auer  and R. Kumar  and J. R. Schmidt  and J. L. Skinner },
title = {Hydrogen bonding and Raman, IR, and 2D-IR spectroscopy of dilute HOD in liquid D<sub>2</sub>O},
journal = {Proceedings of the National Academy of Sciences},
volume = {104},
number = {36},
pages = {14215-14220},
year = {2007},
doi = {10.1073/pnas.0701482104},
URL = {https://www.pnas.org/doi/abs/10.1073/pnas.0701482104},
}

@article{QUIRICO201632,
title = {Refractory and semi-volatile organics at the surface of comet 67P/Churyumov-Gerasimenko: Insights from the VIRTIS/Rosetta imaging spectrometer},
journal = {Icarus},
volume = {272},
pages = {32-47},
year = {2016},
issn = {0019-1035},
doi = {https://doi.org/10.1016/j.icarus.2016.02.028},
url = {https://www.sciencedirect.com/science/article/pii/S001910351600097X},
author = {E. Quirico and L.V. Moroz and B. Schmitt and G. Arnold and M. Faure and P. Beck and L. Bonal and M. Ciarniello and F. Capaccioni and G. Filacchione and S. Erard and C. Leyrat and D. Bockelée-Morvan and A. Zinzi and E. Palomba and P. Drossart and F. Tosi and M.T. Capria and M.C. {De Sanctis} and A. Raponi and S. Fonti and F. Mancarella and V. Orofino and A. Barucci and M.I. Blecka and R. Carlson and D. Despan and A. Faure and S. Fornasier and M.S. Gudipati and A. Longobardo and K. Markus and V. Mennella and F. Merlin and G. Piccioni and B. Rousseau and F. Taylor},
}

@article{MOJIRI2019133971,
title = {Comprehensive review of polycyclic aromatic hydrocarbons in water sources, their effects and treatments},
journal = {Science of The Total Environment},
volume = {696},
pages = {133971},
year = {2019},
issn = {0048-9697},
doi = {https://doi.org/10.1016/j.scitotenv.2019.133971},
url = {https://www.sciencedirect.com/science/article/pii/S0048969719339415},
author = {Amin Mojiri and John L. Zhou and Akiyoshi Ohashi and Noriatsu Ozaki and Tomonori Kindaichi},
}

@article{Sacchi31122023,
author = {M. Sacchi and A. Tamtögl},
title = {Water adsorption and dynamics on graphene and other 2D materials: computational and experimental advances},
journal = {Advances in Physics: X},
volume = {8},
number = {1},
pages = {2134051},
year = {2023},
publisher = {Taylor \& Francis},
doi = {10.1080/23746149.2022.2134051},
URL = {https://doi.org/10.1080/23746149.2022.2134051},
}

@article{Heinrich2,
author = {Salzmann, Heinrich and McCoy, Anne B. and Weber, J. Mathias},
title = {Infrared Spectrum of the Pyrene Anion in the CH Stretching Region},
journal = {The Journal of Physical Chemistry A},
volume = {128},
number = {21},
pages = {4225-4232},
year = {2024},
doi = {10.1021/acs.jpca.4c00966},
note ={PMID: 38753443},
URL = { https://doi.org/10.1021/acs.jpca.4c00966},
}

@book{Farrar_Saunders_1988, place={New York}, title={Techniques for the study of ion-molecule reactions}, publisher={Wiley}, author={Farrar, James M. and Saunders, William Hundley}, year={1988}}

@article{mayo_dreiding_1990, 
  year       = {1990}, 
  title      = {{DREIDING}:  a generic force field for molecular simulations}, 
  author     = {Mayo, Stephen L. and Olafson, Barry D. and Goddard, William A.}, 
  journal    = {The Journal of Physical Chemistry}, 
  issn       = {0022-3654}, 
  doi        = {10.1021/j100389a010}, 
  shorttitle = {{DREIDING}}, 
  pages      = {8897--8909}, 
  number     = {26}, 
  volume     = {94}
}

@article{abascal_general_2005, 
  year       = {2005}, 
  title      = {A general purpose model for the condensed phases of water: {TIP}4P/2005}, 
  author     = {Abascal, J. L. F. and Vega, C.}, 
  journal    = {The Journal of Chemical Physics}, 
  doi        = {10.1063/1.2121687}, 
  shorttitle = {A general purpose model for the condensed phases of water}, 
  pages      = {234505}, 
  number     = {23}, 
  volume     = {123}
}

@article{reimers_local_1984, 
  year    = {1984}, 
  title   = {A local mode potential function for the water molecule}, 
  author  = {Reimers, J.R. and Watts, R.O.}, 
  journal = {Molecular Physics}, 
  issn    = {0026-8976}, 
  doi     = {10.1080/00268978400101271}, 
  pages   = {357--381}, 
  number  = {2}, 
  volume  = {52}
}

@article{singh_approach_1984, 
  year      = {1984}, 
  title     = {An approach to computing electrostatic charges for molecules}, 
  author    = {Singh, U. Chandra and Kollman, Peter A.}, 
  journal   = {Journal of Computational Chemistry}, 
  issn      = {1096-987X}, 
  doi       = {10.1002/jcc.540050204}, 
  pages     = {129--145}, 
  number    = {2}, 
  volume    = {5}
}

@article{besler_atomic_1990, 
  year      = {1990}, 
  title     = {Atomic charges derived from semiempirical methods}, 
  author    = {Besler, Brent H. and Merz, Kenneth M. Jr. and Kollman, Peter A.}, 
  journal   = {Journal of Computational Chemistry}, 
  issn      = {1096-987X}, 
  doi       = {10.1002/jcc.540110404}, 
  pages     = {431--439}, 
  number    = {4}, 
  volume    = {11}
}

@article{batatia2022mace-4b7, 
  year    = {2022}, 
  title   = {{MACE}: Higher Order Equivariant Message Passing Neural Networks for Fast and Accurate Force Fields}, 
  author  = {Batatia, Ilyes and Kovács, Dávid Péter and Simm, Gregor N C and Ortner, Christoph and Csányi, Gábor}, 
  journal = {{arXiv}}, 
  doi     = {10.48550/arxiv.2206.07697}, 
  eprint  = {2206.07697}
}

@article{batatia2022design-dc7, 
  year    = {2022}, 
  title   = {The Design Space of E(3)-Equivariant Atom-Centered Interatomic Potentials}, 
  author  = {Batatia, Ilyes and Batzner, Simon and Kovács, Dávid Péter and Musaelian, Albert and Simm, Gregor N C and Drautz, Ralf and Ortner, Christoph and Kozinsky, Boris and Csányi, Gábor}, 
  journal = {{arXiv}}, 
  doi     = {10.48550/arxiv.2205.06643}, 
  eprint  = {2205.06643}
}

@inbook{zwanzig2001nonequilibrium-8c1, 
  year  = {2001}, 
  author = {Zwanzig, Robert},
  title = {Nonequilibrium Statistical Mechanics}, 
  chapter = {3},
  doi   = {10.1093/oso/9780195140187.001.0001},
  publisher = {Oxford University Press}
}

@article{chance2016advances-f34, 
  year    = {1978}, 
  title   = {Molecular Fluorescence and Energy Transfer Near Interfaces}, 
  author  = {Chance, R. R. and Prock, A. and Silbey, R.}, 
  journal = {Advances in Chemical Physics}, 
  doi     = {10.1002/9780470142561.ch1}, 
  pages   = {1--65},
  volume  = {37}
}

@article{ostrowski2014tunable-303, 
  year    = {2014}, 
  title   = {The Tunable Hydrophobic Effect on Electrically Doped Graphene}, 
  author  = {Ostrowski, Joseph H. J. and Eaves, Joel D.}, 
  journal = {The Journal of Physical Chemistry B}, 
  issn    = {1520-6106}, 
  doi     = {10.1021/jp409342n}, 
  pages   = {530--536}, 
  number  = {2}, 
  volume  = {118}
}

@article{belyaeva2020wettability-186, 
  year    = {2020}, 
  title   = {Wettability of graphene}, 
  author  = {Belyaeva, Liubov A. and Schneider, Grégory F.}, 
  journal = {Surface Science Reports}, 
  issn    = {0167-5729}, 
  doi     = {10.1016/j.surfrep.2020.100482}, 
  pages   = {100482}, 
  number  = {2}, 
  volume  = {75}
}

@article{lopes2004modeling-a0d, 
  year    = {2004}, 
  title   = {Modeling Ionic Liquids Using a Systematic All-Atom Force Field}, 
  author  = {Lopes, José N. Canongia and Deschamps, Johnny and Pádua, Agílio A. H.}, 
  journal = {The Journal of Physical Chemistry B}, 
  issn    = {1520-6106}, 
  doi     = {10.1021/jp0362133}, 
  pages   = {2038--2047}, 
  number  = {6}, 
  volume  = {108}
}

@article{joung2009molecular-797, 
  year    = {2009}, 
  title   = {Molecular Dynamics Simulations of the Dynamic and Energetic Properties of Alkali and Halide Ions Using Water-Model-Specific Ion Parameters}, 
  author  = {Joung, In Suk and Cheatham, Thomas E.}, 
  journal = {The Journal of Physical Chemistry B}, 
  issn    = {1520-6106}, 
  doi     = {10.1021/jp902584c}, 
  pages   = {13279--13290}, 
  number  = {40}, 
  volume  = {113}
}

@article{banerjee2018rotational-0fe, 
  year    = {2018}, 
  title   = {Rotational dynamics of polyatomic ions in aqueous solutions: From continuum model to mode-coupling theory, aided by computer simulations}, 
  author  = {Banerjee, Puja and Bagchi, Biman}, 
  journal = {The Journal of Chemical Physics}, 
  issn    = {0021-9606}, 
  doi     = {10.1063/1.5027031}, 
  pages   = {224504}, 
  number  = {22}, 
  volume  = {148}
}

@article{gastegger2017machine-569, 
  year    = {2017}, 
  title   = {Machine learning molecular dynamics for the simulation of infrared spectra}, 
  author  = {Gastegger, Michael and Behler, Jörg and Marquetand, Philipp}, 
  journal = {Chemical Science}, 
  issn    = {2041-6520}, 
  doi     = {10.1039/c7sc02267k}, 
  pages   = {6924--6935}, 
  number  = {10}, 
  volume  = {8}
}

@article{stocco2025electric-field-2af, 
  year    = {2025}, 
  title   = {Electric-field driven nuclear dynamics of liquids and solids from a multi-valued machine-learned dipolar model}, 
  author  = {Stocco, Elia and Carbogno, Christian and Rossi, Mariana}, 
  journal = {npj Computational Materials}, 
  doi     = {10.1038/s41524-025-01751-x}, 
  pages   = {304}, 
  number  = {1}, 
  volume  = {11}
}

@article{behler2014representing-f36, 
  year    = {2014}, 
  title   = {Representing potential energy surfaces by high-dimensional neural network potentials}, 
  author  = {Behler, J}, 
  journal = {Journal of Physics: Condensed Matter}, 
  issn    = {0953-8984}, 
  doi     = {10.1088/0953-8984/26/18/183001}, 
  pages   = {183001}, 
  number  = {18}, 
  volume  = {26}
}

@article{bartk2017machine-0f1, 
  year    = {2017}, 
  title   = {Machine learning unifies the modeling of materials and molecules}, 
  author  = {Bartók, Albert P. and De, Sandip and Poelking, Carl and Bernstein, Noam and Kermode, James R. and Csányi, Gábor and Ceriotti, Michele}, 
  journal = {Science Advances}, 
  doi     = {10.1126/sciadv.1701816}, 
  pages   = {e1701816}, 
  number  = {12}, 
  volume  = {3}
}

@article{kovcs2023evaluation-a40, 
  year    = {2023}, 
  title   = {Evaluation of the {MACE} force field architecture: From medicinal chemistry to materials science}, 
  author  = {Kovács, Dávid Péter and Batatia, Ilyes and Arany, Eszter Sára and Csányi, Gábor}, 
  journal = {The Journal of Chemical Physics}, 
  issn    = {0021-9606}, 
  doi     = {10.1063/5.0155322}, 
  pages   = {044118}, 
  number  = {4}, 
  volume  = {159}
}

@article{bhatia2025leveraging-07f, 
  year    = {2025}, 
  title   = {Leveraging active learning-enhanced machine-learned interatomic potential for efficient infrared spectra prediction}, 
  author  = {Bhatia, Nitik and Rinke, Patrick and Krejčí, Ondřej}, 
  journal = {npj Computational Materials}, 
  doi     = {10.1038/s41524-025-01827-8}, 
  pages   = {324}, 
  number  = {1}, 
  volume  = {11}
}

@article{schran2021machine-c45, 
  year    = {2021}, 
  title   = {Machine learning potentials for complex aqueous systems made simple}, 
  author  = {Schran, Christoph and Thiemann, Fabian L. and Rowe, Patrick and Müller, Erich A. and Marsalek, Ondrej and Michaelides, Angelos}, 
  journal = {Proceedings of the National Academy of Sciences}, 
  issn    = {0027-8424}, 
  doi     = {10.1073/pnas.2110077118},  
  pages   = {e2110077118}, 
  number  = {38}, 
  volume  = {118}
}

@article{eltareb2025understanding-12a, 
  year    = {2025}, 
  title   = {Understanding Isotope Substitution Effects in Water Using the Potential Energy Landscape Formalism for Quantum Liquids}, 
  author  = {Eltareb, Ali and Zhou, Yang and Lopez, Gustavo E. and Giovambattista, Nicolas}, 
  journal = {Journal of Chemical Theory and Computation}, 
  issn    = {1549-9618}, 
  doi     = {10.1021/acs.jctc.5c01325}, 
  pages   = {11931--11950}, 
  number  = {23}, 
  volume  = {21}
}

@article{feller1999strength-bdb, 
  year    = {1999}, 
  title   = {Strength of the Benzene--Water Hydrogen Bond}, 
  author  = {Feller, David}, 
  journal = {The Journal of Physical Chemistry A}, 
  issn    = {1089-5639}, 
  doi     = {10.1021/jp991932w}, 
  pages   = {7558--7561}, 
  number  = {38}, 
  volume  = {103}
}

@article{gierszal2011-hydrogen-71f, 
  year    = {2011}, 
  title   = {$\pi-$Hydrogen Bonding in Liquid Water}, 
  author  = {Gierszal, Kamil P. and Davis, Joel G. and Hands, Michael D. and Wilcox, David S. and Slipchenko, Lyudmila V. and Ben-Amotz, Dor}, 
  journal = {The Journal of Physical Chemistry Letters}, 
  issn    = {1948-7185}, 
  doi     = {10.1021/jz201373e}, 
  pages   = {2930--2933}, 
  number  = {22}, 
  volume  = {2}
}

@article{adekola2019effects-de3, 
  year    = {2019}, 
  title   = {Effects of ground plane dimensions on patterns radiated by slot antennas}, 
  author  = {Adekola, S. Adeniyi and Amusa, K. Akinwale}, 
  journal = {International Journal of Electronics}, 
  issn    = {0020-7217}, 
  doi     = {10.1080/00207217.2019.1584922}, 
  pages   = {1295--1319}, 
  number  = {9}, 
  volume  = {106}
}

@article{galabov2002high-b9d, 
  year    = {2002}, 
  title   = {High Level ab Initio Quantum Mechanical Predictions of Infrared Intensities}, 
  author  = {Galabov, Boris and Yamaguchi, Yukio and Remington, Richard B. and Schaefer, Henry F.}, 
  journal = {The Journal of Physical Chemistry A}, 
  issn    = {1089-5639}, 
  doi     = {10.1021/jp013297b}, 
  pages   = {819--832}, 
  number  = {5}, 
  volume  = {106}
}

@article{mackerell1998all-atom-e5f, 
  year    = {1998}, 
  title   = {All-Atom Empirical Potential for Molecular Modeling and Dynamics Studies of Proteins †}, 
  author  = {{MacKerell}, A. D. and Bashford, D. and Bellott, M. and Dunbrack, R. L. and Evanseck, J. D. and Field, M. J. and Fischer, S. and Gao, J. and Guo, H. and Ha, S. and Joseph-{McCarthy}, D. and Kuchnir, L. and Kuczera, K. and Lau, F. T. K. and Mattos, C. and Michnick, S. and Ngo, T. and Nguyen, D. T. and Prodhom, B. and Reiher, W. E. and Roux, B. and Schlenkrich, M. and Smith, J. C. and Stote, R. and Straub, J. and Watanabe, M. and Wiórkiewicz-Kuczera, J. and Yin, D. and Karplus, M.}, 
  journal = {The Journal of Physical Chemistry B}, 
  issn    = {1520-6106}, 
  doi     = {10.1021/jp973084f}, 
  pmid    = {24889800}, 
  pages   = {3586--3616}, 
  number  = {18}, 
  volume  = {102}
}

@article{suzuki1992benzene-d1b, 
  year    = {1992}, 
  title   = {Benzene Forms Hydrogen Bonds with Water}, 
  author  = {Suzuki, Sakae and Green, Peter G. and Bumgarner, Roger E. and Dasgupta, Siddharth and {III}, William A. Goddard and Blake, Geoffrey A.}, 
  journal = {Science}, 
  issn    = {0036-8075}, 
  doi     = {10.1126/science.257.5072.942}, 
  pmid    = {17789637}, 
  pages   = {942--945}, 
  number  = {5072}, 
  volume  = {257}
}

@article{schravendijk2005from-7e5, 
  year    = {2005}, 
  title   = {From Hydrophobic to Hydrophilic Solvation: An Application to Hydration of Benzene}, 
  author  = {Schravendijk, Pim and Vegt, Nico F. A. van der}, 
  journal = {Journal of Chemical Theory and Computation}, 
  issn    = {1549-9618}, 
  doi     = {10.1021/ct049841c}, 
  pmid    = {26641686}, 
  pages   = {643--652}, 
  number  = {4}, 
  volume  = {1}
}

@article{allesch2008first-651, 
  year    = {2008}, 
  title   = {First principles and classical molecular dynamics simulations of solvated benzene}, 
  author  = {Allesch, Markus and Lightstone, Felice C. and Schwegler, Eric and Galli, Giulia}, 
  journal = {The Journal of Chemical Physics}, 
  issn    = {0021-9606}, 
  doi     = {10.1063/1.2806288}, 
  pmid    = {18190198}, 
  pages   = {014501}, 
  number  = {1}, 
  volume  = {128}
}

\newpage

\renewcommand{\thefigure}{S\arabic{figure}}
\renewcommand{\thetable}{S\arabic{table}}
\renewcommand{\theequation}{S\arabic{equation}}
\renewcommand{\thepage}{S\arabic{page}}
\setcounter{figure}{0}
\setcounter{table}{0}
\setcounter{equation}{0}
\setcounter{page}{1}

\newpage

\clearpage

\end{document}